\documentclass{workshop}
\usepackage{graphicx}
\begin{document}

\title{
Bayesian Time-Resolved Spectroscopy of GRB Pulses: \\$\alpha$-Intensity Correlation
}

\author{
H\"usne Dereli B\'egu\'e, $^{1,2,3}$Hoi-Fung Yu,$^{2,3,4}$ Felix Ryde,$^{2,3}$ \\ 
\\[12pt]  
%
$^{1}$ Max Planck Institute for Extraterrestrial Physics, Giessenbachstrasse 1, D-85748 Garching, Germany \\
$^{2}$ Department of Physics, KTH Royal Institute of Technology, AlbaNova, SE-10691, Stockholm, Sweden \\
$^{3}$ The Oskar Klein Centre for Cosmoparticle Physics, SE-10691, Stockholm, Sweden \\
$^{4}$ Faculty of Science, The University of Hong Kong, Pokfulam, Hong Kong  \\
%
\textit{E-mail: husnedereli@gmail.com} 
}

\abst{
Gamma-ray bursts (GRBs) show different behaviours and trends in their spectral evolution. One of the methods used to understand the physical origin of these behaviours is to study correlation between the spectral fit parameters. In this work, we used a Bayesian analysis method to fit time-resolved spectra of GRB pulses that were detected by the \textit{Fermi}/GBM during its first 9 years of mission. We studied single pulsed long bursts ($T_{90}\geq2$ s). Among all the parameter correlations, we found that the correlation between the low-energy power-law index $\alpha$ and the energy flux exhibited a systematic behaviour. We presented the properties of the observed characteristics of this behaviour and interpreted it in the context of the photospheric emission model.
}

\kword{catalogs - gamma-ray burst: general - methods: data analysis and statistical}
\maketitle
\thispagestyle{empty}

\section{Introduction}
The prompt phase of GRBs is characterized by a high flux of gamma-ray photons (keV - MeV). In this phase, each source shows a different behavior and trend in its light-curve. These features are confronted by different models (e.g. synchrotron, photospheric models) in order to be explained.

In \citet{Yu2019},  we studied a sample of 38 single pulses from 37 long gamma-ray bursts ($T_{90}\geq2$ s) detected by the \textit{Fermi}/Gamma-ray Burst Monitor (GBM) during its first 9 years of mission. We performed a temporal binning to the data via Bayesian block \citep{Scargle2013}. We required a statistical significance of $S \geq 20$ for each bin \citep{Vianello2015}. We then performed a time-resolved spectroscopy in each time bin individually and analysed a total of 577 spectra. We fitted both a BAND function and a cut off power law (CPL) function.
We found that the CPL model can describe most of the spectra. Therefore, we mainly studied the time evolution of the spectral parameters and their relation within each pulse individually using the CPL model.

\section{Parameter Distributions}
The distributions of the low-energy power-law index ($\alpha$) and the peak energy of the $\nu F_{\nu}$ spectra ($E_{\mathrm p}$) are in agreement with previous time-resolved catalogs, e.g. \citet{Yu2016} which though include not only individual pulses. Therefore, the frequentist and Bayesian analyses give consistent results. However, the distribution of the high-energy power-law index ($\beta$) has lower values than previously found. This indicates that single pulses are in general softer.

\begin{figure}[hbt]
\centering
\includegraphics[width=\linewidth]{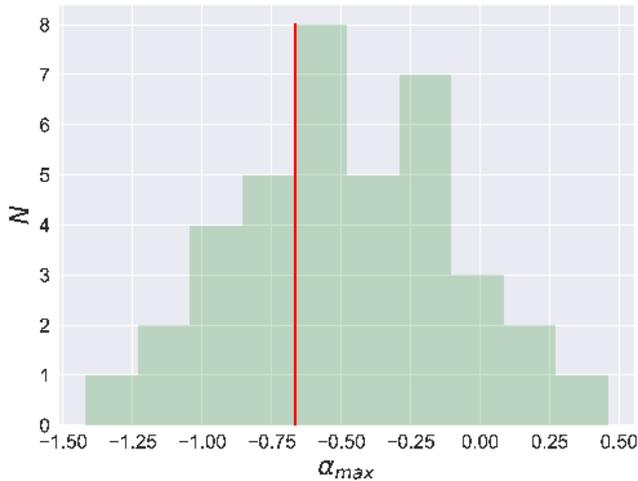}
\caption{Histogram of the maximal value of $\alpha$ in each of the 38 pulses in the sample. The red line indicate the synchrotron "line of death" ($-2/3$). All values on the right of this line are incompatible with synchotron radiation. This Figure is taken from \citet{Yu2019}.
\label{fig:alpha_max}}
\end{figure}

We also studied the $\alpha_{\mathrm{max}}$ distribution: we selected the maximal (hardest) value of $\alpha$ in each pulse. Under the assumption that a single emission mechanism is responsible for the full duration of the pulse emission such a distribution is indicative of the emission during the whole pulse. We found that  $60\%$ of pulses (with a $1\sigma$ lower error limit) have values of $\alpha_{\mathrm{max}}$ which are larger than $-2/3$ and violate the criteria set by the "synchrotron line of death" (see Figure \ref{fig:alpha_max}). This fraction is significantly larger than what is found before ($\sim28\%$, in \cite{Preece1998}).

\section{The temporal parameter evolution}
We found that $\alpha_{\mathrm{CPL}}$ and $\alpha_{\mathrm{BAND}}$ track each other as well as the flux in most pulses. However, $\alpha_{\mathrm{BAND}}$ tends to have slightly higher values than $\alpha_{\mathrm{CPL}}$. The $\beta_{\mathrm{BAND}}$ is usually smaller than $-3$. 
We  observed that the evolution of $E_{\mathrm{p}}$ exhibits various trends: most cases (16 cases; $42\%$) exhibit pure hard-to-soft evolution,  while 8 cases ($21\%$) exhibit pure intensity tracking evolution. However, $E_{\mathrm{p,BAND}}$ is always smaller than $E_{\mathrm{p,CPL}}$. 
We showed that the calculated energy fluxes ($F$) from the CPL and the BAND model fit parameters track the count light curve. However, during low-significance time bins: $F_{\mathrm{BAND}}$ is always larger than $F_{\mathrm{CPL}}$.

\section{The parameter relations}
We studied the parameter relations between the fit parameters, $E_{\mathrm{p}}$, $F$ and $\alpha$.
The $\alpha-E_{\mathrm{p}}$ relations show three main types of behaviours: 17 pulses show non-monotonic relation, with a clear break, 12 pulses (7 positive and 5 negative) show monotonic relation, straight line in the linear-log plots, In 7 pulses, $E_{\mathrm{p}}$ does not vary much, while $\alpha$ does vary more significantly. 

The $F-E_{\mathrm{p}}$ relations also show three main types of behaviours: 23 pulses show non-monotonic relation, with a distinct break and having power-law segments, 13 pulses (11 positive and 2 negative) has a single power law, 2 pulses do not have any clear trend.

In contrast to these two relations, the $F-\alpha$ relations have a more homogeneous behaviour: 34 pulses have a positive and linear relation (in the semi-log plots). To quantify the observed correlations, the Spearman’s rank coefficient ($r$) was calculated; 28 pulses have $r > 0.7$, indicating a strong correlation, 8 pulses even have very strong correlations ($r > 0.9$), while only two pulses have weak correlations ($r < 0.4$).

\section{The $\alpha$-intensity correlation}
It is natural to assume a photospheric emission model for the burst with $-2/3<\alpha_{\mathrm{max}}<0$ in Figure \ref{fig:alpha_max}. A qualitative photospheric emission scenario is presented in \citet{Ryde2017}. In this scenario, there are two expectations. The first one is that an intense emission with a narrow spectrum is expected because around the peak of the light curve a large entropy causes the photosphere to approach the saturation radius. The second one is that a weaker and softer emission is expected when the entropy decreases, the photosphere is at larger radii than the saturation radius. This scenario naturally leads to a correlated variation of the intensity and spectral shape, covering the observed range. 

We then performed a linear fit between $\alpha$ and the logarithm of the flux, characterised by the slope $k$. The values of $k$ ranges from 1 to 5, with a mean of 2.80 for CPL and 3.67 for BAND.

The slope $k$ is consistent with expectations ($\sim$ 3) from the photospheric emission scenario if the flux variation is due to adiabatic cooling. We find that $70\%$ of the long bursts are consistent with $k =3$.

\section{Conclusion}
In \citet{Yu2019}, we showed that more than half of the pulses are inconsistent with synchrotron emission, solely based on the "line of death" of $\alpha = -2/3$. This suggest that the emission mechanism is different. In addition, we found that the fitted parameter relations show a variety of behaviours, however, the $\alpha$-intensity correlation is the most homogenious. We explained this correlation within a qualitative photospheric emission scenario, in an outflow where the dimensionless entropy varies. We found that $70\%$ of the long bursts are consistent with this scenario \citep{Ryde2019}. 

For further use, the analysis result files in FITS format for every time bin are available at https://zenodo.org/record/2601901

\label{last}

\end{document}